\documentclass[%
 reprint,
 amsmath,amssymb,
 aps,
]{revtex4-2}
\usepackage{dsfont}
\usepackage{bbm}
\usepackage{graphicx}% Include figure files
\usepackage{dcolumn}% Align table columns on decimal point
\usepackage{bm}% bold math
\usepackage{hyperref}% add hypertext capabilities
\usepackage{color}

\def \e{{\mathbf e}}

\def \k{{\mathbf k}}
\def \q{{\mathbf q}}

\def \r{{\mathbf r}}

\def \e{{\mathbf e}}

\def \n{{\mathbf n}}
\def \d{{\mathbf d}}
\def \r{{\mathbf r}}

\def \Q{{\mathbf Q}}

\begin{document}

\preprint{APS/123-QED}

\title{Superconductivity from emergent dipolar interactions in a fractionalized Fermi liquid}

\author{Mina-Lou Schleith}
 \affiliation{Department of Physics and Astronomy, Ghent University, Krijgslaan 299,  9000 Ghent, Belgium}
\author{Arthur Bril}
\affiliation{Department of Physics and Astronomy, Ghent University, Krijgslaan 299,  9000 Ghent, Belgium}
\author{Nick Bultinck}
 \affiliation{Department of Physics and Astronomy, Ghent University, Krijgslaan 299,  9000 Ghent, Belgium}

\date{\today}

\begin{abstract}
Starting from the spin-fermion model or Hertz-Millis theory describing electrons coupled to anti-ferromagnetic spin fluctuations we develop a theory to describe the transition from a fractionalized Fermi liquid into a $d_{x^2-y^2}$ superconductor. We focus on small electron doping on top of the half-filled state. The doped electrons enter the system as spinon-chargon bound states, which form a small, reconstructed Fermi surface. The bound states are neutral under the emergent U(1) gauge symmetry of the fractionalized Fermi liquid, but interact via a dipolar two-body potential. We show that because of the projective action of translation symmetry on the spinons and chargons, the Fourier components of this repulsive dipolar interaction are peaked at the anti-ferromagnetic wave vector, thereby providing a robust microscopic mechanism for $d_{x^2-y^2}$ pairing in a fractionalized metal.
\end{abstract}

%\keywords{Suggested keywords}%Use showkeys class option if keyword
                              %display desired
\maketitle

%\tableofcontents
\section{Introduction}

Fractionalized Fermi liquids (FL$^*$) are metallic phases with Fermi surfaces of electron-like or hole-like quasi-particles enclosing a volume that violates Luttinger's theorem due to the presence of a non-trivial topological order~\cite{FLstar}. As the name suggests, FL$^*$ phases also host fractionalized excitations such as spinons and chargons, and sometimes the electron-like and hole-like quasiparticles can be interpreted as bound states of the former~\cite{Kaul_2007,Punk2015,Sachdev2016}. The perspective that holes should be thought of as spinon-chargon bound states has a long history in the study of the cuprate superconductors \cite{Beran1996,Baskaran}, where it is also sometimes expressed in a flipped-spin string picture of a hole moving in an anti-ferromagnetic background \cite{Brinkman,Trugman,Shraiman_1988_HeisenbergAFM,Dagotto,Grusdt2018,Christie2019,Koepsell2019}.

\begin{figure}[t!]
    \centering
    \includegraphics[width=0.8\linewidth] {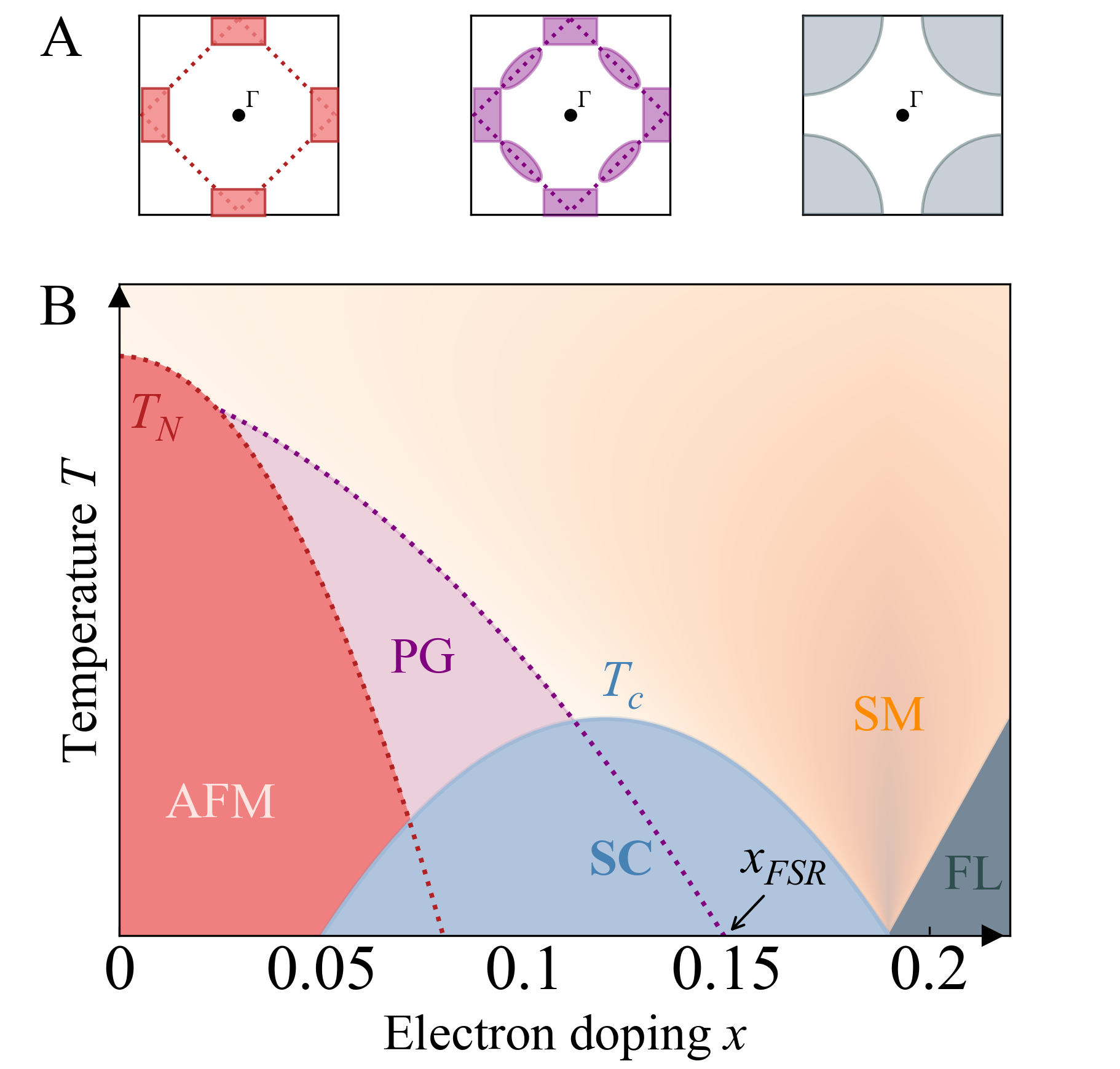}
    \caption{\textbf{A:} Evolution of the Fermi surface with doping in the electron-doped cuprates. \textbf{B:} Schematic of the electron-doped cuprate phase diagram. We assume that in the doping range between the offset of anti-ferromagnetic (AFM) order and the Fermi surface reconstruction at $x_{FSR}$ (i.e. the `pseudogap (PG)' region) the normal state of the superconductor (SC) is a fractionalized Fermi liquid.} 
    \label{fig:cuprate_phase_diagram}
\end{figure}

Despite the spinon-chargon bound state perspective being fairly well-established, there seems to be little understanding of how these bound states would form Cooper pairs -- clearly a central question in the context of the cuprate superconductors. In this work we address this question, having the electron-doped cuprates in mind. There is experimental evidence that the electron-doped compounds also exhibit Fermi surface reconstruction in the absence of symmetry breaking~\cite{Helm2010,He2019,Xu2023} (for a schematic of the phase diagram, see Fig. \ref{fig:cuprate_phase_diagram}), and they provide the conceptual simplification of only having to consider anti-ferromagnetic spin fluctuations, as stripe and charge density wave orders are absent in the phase diagram. We will therefore start our analysis from the Hertz-Millis theory for fluctuating itinerant anti-ferromagnetism (also called the spin-fermion model in the high-$T_c$ context), and introduce fractionalization by going to the rotating spin reference frame~\cite{Shraiman_1988_HeisenbergAFM,Schulz,Dupuis,Borejsza,Kaul_2007,Kaul2008,Sachdev_fluctuating_2009,Qi2010,Chowdhury2015,Sachdev2016,Chatterjee2017,Chatterjee2017_2,Scheurer2018,Wu2018,Sachdev2019,SachdevScammell,Zhang2020,Bonetti,Stepanov2022,Nikolaenko2023,Goremykin2024,Nikolaenko,Vilardi,Forni,Sachdev2025,Muller2026}.

Our main result is that the projective action of translation on the spinons and chargons in the FL$^*$ state has important consequences for the interaction between spinon-chargon bound states. In particular, the FL$^*$ theory used in this work has a U(1) gauge field which generates a dipolar interaction between the gauge-neutral spinon-chargon bound states. The projective symmetry action endows this dipolar interaction with a very unconventional property: in momentum space it is peaked not at zero momentum, but at the anti-ferromagnetic wave vector. This is generally expected to lead to $d_{x^2-y^2}$-wave superconductivity, as is confirmed by our numerical solution of the Bardeen-Cooper-Schrieffer (BCS) gap equation. We thus find that non-trivial projective symmetry actions associated with fractionalization provide a microscopic mechanism for unconventional superconductivity in the BCS regime.

\section{Rotating reference frame theory} 

In contrast to the hole-doped cuprates, the physics of the electron-doped cuprates seems to be completely determined by anti-ferromagnetic (AFM) spin fluctuations. We therefore start from an effective theory with Euclidean action $S=S_b + S_f$, where $S_b$ is a non-linear sigma model describing the AFM fluctuations:
\begin{equation}
S_b = \frac{1}{4} \int_0^\beta\mathrm{d}\tau \left[\sum_\r \chi (\partial_\tau \n_\r)^2 - J \sum_{\langle \r,\r'\rangle} \n_\r\cdot \n_{\r'} \right].  \label{Sb}
\end{equation}
Here $\n_\r$ is a three-component unit vector field, and $\chi,J>0$. The spin fluctuations are coupled to the electrons, which are described by $S_f$:
\begin{eqnarray}
S_f & = & \int_0^\beta \mathrm{d}\tau \bigg[ \sum_\r \bar{c}_\r(\partial_\tau -\mu) c_\r + g (-1)^{x+y}\, \n_\r \cdot \bar{c}_\r \boldsymbol{\sigma} c_\r \nonumber \\
& &  - t \sum_{\langle \r,\r'\rangle} (\bar{c}_\r c_{\r'} +h.c.)- t'\sum_{\langle\langle \r,\r'\rangle\rangle}(\bar{c}_\r c_{\r'} +h.c.) \bigg]. 
\label{Sf}
\end{eqnarray}
In the fermion part of the action we have introduced a chemical potential $\mu$, a coupling strength $g$, and nearest and next-nearest hopping terms with respective amplitudes $t$ and $t'$. Note that the coupling term contains a sign factor which ensures that the bosonic field indeed induces AFM fluctuations for the fermion spins. Because of this staggered coupling, the translation symmetry acts on the fields as
\begin{eqnarray}
c_\r & \rightarrow & c_{\r + \e_{x/y}}, \\
\n_\r & \rightarrow & - \n_{\r + \e_{x/y}}. \label{minn}
\end{eqnarray}
Note in particular the minus sign that $\n_\r$ acquires under a single-site translation -- this will play a crucial role for our main results presented below.

We assume that the pseudogap region of the electron-doped phase diagram corresponds to a fractionalized Fermi liquid, where Luttinger's theorem is violated due to the presence of non-trivial topological order. To describe this topological metal we first fractionalize the bosonic field $\n_\r$ in terms of a spin-1/2 spinon field $z_\r$: 
\begin{equation}
\n_\r = z^*_\r \boldsymbol{\sigma} z_\r.
\end{equation}
This parameterization of $\n_\r$ introduces a U(1) gauge freedom, as $\n_\r$ is invariant under changes in the overall phase of $z_\r$. From Eq. \eqref{minn} we see that the spinon field transforms under a single-site translation as
\begin{equation}
z_\r \rightarrow \mathrm{i}\sigma^y z_{\r+\e_{x/y}}^*.
\end{equation}
Crucially, this transformation property implies that U(1) gauge charges change sign under elementary translations.

From the spinon field we can define an SU(2) matrix field
\begin{equation}
    R_{\bm r}=
    \begin{pmatrix} 
        z_{\bm r,\uparrow} & -z_{\bm r,\downarrow}^* \\ z_{\bm r,\downarrow} & z_{\bm r,\uparrow}^*
    \end{pmatrix} \, ,
\end{equation}
which has following property:
\begin{equation}
    \bm n_{\bm r} \cdot \bm \sigma = R_{\bm r} \sigma^z R_{\bm r}^\dagger.
    \label{eq:ndotsigma}
\end{equation}
Using $R_\r$ we in turn define new fermion fields $\psi_\r$, $\bar{\psi}_\r$, which live a rotating spin reference frame~\cite{Shraiman_1988_HeisenbergAFM,Schulz,Dupuis,Borejsza,Kaul_2007,Kaul2008,Sachdev_fluctuating_2009,Qi2010,Chowdhury2015,Sachdev2016,Chatterjee2017,Chatterjee2017_2,Scheurer2018,Wu2018,Sachdev2019,SachdevScammell,Zhang2020,Bonetti,Stepanov2022,Nikolaenko2023,Goremykin2024,Nikolaenko,Vilardi,Forni,Sachdev2025,Muller2026}:
\begin{eqnarray}
\bar{\psi}_\r & = & \bar{c}_\r R_\r, \\
\psi_\r & = & R_\r^\dagger c_\r.
\end{eqnarray}
The new fermion fields $\bar{\psi}_{\r,+} = \bar{c}_\r z_\r$ and $\bar{\psi}_{\r,-} = -\bar{c}_\r \mathrm{i}\sigma^y z^*_\r$ respectively have U(1) gauge charge $+1$ and $-1$, and are spin singlet. This allows us to identify them as the chargon excitations.

We can now exactly rewrite $S_f$ in terms of the chargon fields:
\begin{eqnarray}
S_f & = & \int\mathrm{d}\tau \bigg[ \sum_{\r,\r'} \bar{\psi}_\r(\delta_{\r,\r'} (\partial_\tau-\mu)  + t_{\r\r'} R^\dagger_\r R_{\r'}) \psi_{\r'}  \\
&& \hspace{0.2 cm} + \sum_\r \bar{\psi}_\r(R_\r^\dagger\partial_\tau R_\r)\psi_\r 
 + g\sum_\r (-1)^{x+y} \bar{\psi}_\r \sigma^z \psi_\r \bigg]. \nonumber
\end{eqnarray}
Note in particular that because of Eq. \eqref{eq:ndotsigma} the electron-boson coupling term has become a static staggered potential for the chargons, which ensures that the free chargon dispersion is gapped at half-filling. This staggered potential is invariant under translations, which act as $\psi_\r \rightarrow \mathrm{i}\sigma^y \psi_{\r+\e_{x/y}}$.

To make the physical interpretation of the other terms in $S_f$ more clear we split the action as $S_f = S_\psi + S_{\psi z}$, with
\begin{eqnarray}
S_\psi & = & \int \mathrm{d}\tau \bigg[ \sum_{\r,\r'} \bar{\psi}_\r(\delta_{\r,\r'} \left[\partial_\tau -\mu +  \mathrm{i}a_{0} \sigma^z \right] + \tilde{t}_{\r\r'} e^{\mathrm{i}a_{\r\r'}\sigma^z}) \psi_{\r'} \nonumber \\
&& \hspace{2 cm} 
 + g\sum_\r (-1)^{x+y} \bar{\psi}_\r \sigma^z \psi_\r \bigg] \label{Spsi}\, ,
\end{eqnarray}
Here $\tilde{t}_{\r\r'} = |z^*_\r z_{\r'}| t_{\r\r'}$ will be interpreted as effective nearest and next-nearest neighbour hopping parameters, and $a_0 = -\mathrm{i}z^*\partial_\tau z$, $a_{\r\r'} = \text{arg}(z^*_\r z_{\r'})$ are the temporal and spatial components of a U(1) lattice gauge field. The other contribution to $S_f$ takes the form
\begin{eqnarray}
S_{\psi z} & = & \int\mathrm{d}\tau \bigg[\sum_{\r,\r'} t_{\r\r'} \bar{\psi}_{\r,-} \psi_{\r',+}\, z_{\r'}\mathrm{i}\sigma^y z_\r + \text{h.c.} \label{SrSi}\\
 & & \hspace{2 cm} + \sum_\r \bar{\psi}_{\r,-} \psi_{\r,+}\, z_{\r}\mathrm{i}\sigma^y \partial_\tau z_\r + \text{h.c.}  \bigg]. \nonumber
\end{eqnarray}
These are the Shraiman-Siggia terms~\cite{Shraiman_1988_HeisenbergAFM}, which represent a local spinon-chargon interaction.

The most straightforward way to express the bosonic part of the action in terms of the spinon fields is via a continuum approximation, as this allows us to exactly rewrite it as
\begin{eqnarray}
S_{b,\text{cont}} & =  & \frac{\chi}{4} \int_0^\beta \mathrm{d}\tau \int\mathrm{d}^2\r\, (\partial_\mu \n)^2 \\
 &= & \chi \int_0^\beta \mathrm{d}\tau \int \mathrm{d}^2\r\, |(\partial_\mu - ia_\mu) z|^2\,,
\end{eqnarray}
where $\partial_\mu = (\partial_\tau,v\boldsymbol{\nabla})$ with $v = \sqrt{J/2\chi}$. At the saddle-point level, $a_\mu$ can be identified as
\begin{equation}
a_\mu = -\mathrm{i}z^* \partial_\mu z\,,
\end{equation}
which is the continuum version of the U(1) gauge field introduced previously in Eq. \eqref{Spsi} (up to a factor of $v$ for the spatial components).

The fractionalized Fermi liquid that we consider in this work is realized by taking the spinons to be gapped, and the U(1) gauge field to be in its deconfined phase.

\section{Spinon-chargon bound states}
We now consider the half-filled state with gapped spinons and chargons, and imagine injecting a single electron into the system. This will result in the simultaneous creation of a spinon and chargon, which interact strongly via the U(1) gauge field. Our goal is to, within certain approximations, study the bound states formed by this spinon-chargon pair. We will reformulate the problem in the Hamiltonian formalism, where the effect of the U(1) gauge field is to introduce a Coulomb interaction between the chargon and spinon. For this first step we will closely follow the analysis of Ref.~\cite{Kaul_2007}. 

\subsection{Coulomb interaction}
The action contains no bare kinetic term for the U(1) gauge field. To introduce dynamics for the gauge field, we assume that the spinon mass is significantly smaller than the chargon mass, and use the spinon polarization to generate a dynamical term
\begin{eqnarray}
    S_a = \frac{1}{2}\sum_{q} a_\mu (q)\Pi(q)  \left[\delta_{\mu\nu} - \frac{q_\mu q_\nu}{q^2} \right]a_\nu(-q).
    \label{eq:effectivephotonaction}
\end{eqnarray} 
Here we have introduced the notation $q=(\Omega,v\q)$. To get an expression for $\Pi(q)$ we start from the continuum action in Eq. \eqref{eq:effectivephotonaction} and do the standard saddle-point analysis~\cite{Polyakov_1987}, i.e. impose $|z|^2=1$ via a Lagrange multiplier, and then fix the Lagrange multiplier to its saddle-point value to generate a non-zero mass $m$ for the spinons. Next one can calculate the standard one-loop paramagnetic and diamagnetic contributions to the spinon polarization to find~\cite{Kaul_2007}
\begin{equation}
    \Pi(q) = \frac{q^2+4m^2}{8\pi v^2|q|}\arctan{\frac{|q|}{2m}} - \frac{m}{4\pi v^2},
    \label{eq:photon_propagator}
\end{equation}
To proceed we will work in the Coulomb gauge, and only consider the density-density interaction mediated by $a_0$. The current-current interaction mediated by the transverse photon will be ignored. As $\Pi(q) = q^2/16\pi v^2 m $ for $|q|\ll m$, the $a_0$ propagator becomes frequency independent at low energies and momenta. We will study the bound state problem in the Hamiltonian formalism, for which we replace the complete polarization (so not only for small $|q|$) by a frequency-independent expression by setting $\Omega=0$, such that it realizes an instantaneous interaction, which is the Coulomb interaction for this system. 

Next we introduce $\pi$, the field canonically conjugate to $z$, and the mode decompositions
\begin{eqnarray}
z_{\q} & = & \frac{1}{\sqrt{2\omega_{\q}\chi}}(b_{\q}+\mathrm{i}\sigma^y\bar{a}_{-\q}), \label{mode1} \\
\pi_\q & = & \mathrm{i}\sqrt{\frac{\omega_\q \chi }{2}} (b_\q - \mathrm{i}\sigma^y\bar{a}_{-\q})\,, \label{mode2}
\end{eqnarray}
where $\omega_\q=\sqrt{(v\q)^2+m^2}$ is the spinon dispersion. Ignoring the transverse gauge field component, the spinon action can be written in terms of the spinon mode operators as
\begin{eqnarray}
S_{b} & = & \int_0^\beta\mathrm{d}\tau \sum_\q \bigg[  \bar{b}_\q \partial_\tau b_\q + \bar{a}_\q \partial_\tau a_\q + \omega_\q (\bar{a}_\q a_\q + \bar{b}_\q b_\q)  \nonumber \\
& & \hspace{0.5 cm} + \frac{\mathrm{i}}{\sqrt{N}}\sum_{\q'}a_0(\q')(\bar{a}_{\q}a_{\q+\q'} - \bar{b}_\q b_{\q+\q'} ) \bigg] \,,
\end{eqnarray}
where $N$ is the number of sites in the system. After bringing the action in this form, it is clear that integrating out $a_0$ produces a Coulomb interaction
\begin{equation}
H_C = \frac{1}{2} \sum_{\q} V_\q \hat \rho_\q \hat \rho_{-\q}\,,
\end{equation}
where the interaction potential is given by
\begin{equation}
V_\q = \Pi(0,v\q)^{-1} = \left[\frac{v^2\q^2+4m^2}{8\pi v^3|\q| } \arctan\frac{v|\q|}{2m} - \frac{m}{4\pi v^2}\right]^{-1},
\end{equation}
and $\rho_\q$ is the gauge charge density:
\begin{equation}
\hat \rho_\q = \frac{1}{\sqrt{N}}\sum_{\q'} \left[\hat \psi^\dagger_{\q}\sigma^z \hat \psi_{\q+\q'}  + \hat a^\dagger_{\q} \hat a_{\q+\q'} - \hat b^\dagger_\q \hat b_{\q+\q'} \right].
\end{equation}
As the interaction potential was obtained using a continuum approximation for the spinons, we need to restore the periodicity under shifts by reciprocal lattice vectors. We do this via the replacement $|\bm q| \to 2\sqrt{\sin^2(q_x/2)+\sin^2(q_y/2)}$. We also add a momentum-independent shift to the potential to tune the on-site part of the interaction, which is not expected to be faithfully described by the long-wavelength effective theory. Even though we used many approximations to get at the final Coulomb potential, its precise form will actually not matter for the main results presented below, as they are all fixed by the projective symmetry action of the problem. We nevertheless use the spinon polarization in order to get (1) a realistic order of magnitude for the interaction, (2) the correct trends in the parametric dependence on $v$ and $m$, and (3) the characteristic $1/\q^2$ divergence at small $\q$.

\subsection{Shraiman-Siggia interaction}
The Shraiman-Siggia contribution to the action in Eq.~\eqref{SrSi} leads to a spinon-chargon interaction of the form
\begin{align}
\hat H_{\psi z} & =  \frac{1}{N} \sum_{\bm k\bm k'\bm q} \bigg[\frac{\varepsilon_{\bm k'+\bm k-\bm q}-\varepsilon_\q}{2\chi \sqrt{\omega_{-\bm k+\bm q}\omega_{-\bm k'+\bm q}}} \times \\ & (\hat \psi_{+,\k}^\dagger \hat b^\dagger_{-\bm k+\bm q} \hat \psi_{-,\bm k'} \hat a_{-\bm k'+\bm q} 
    + \hat \psi_{-,\k}^\dagger \hat a^\dagger_{-\bm k+\bm q} \hat \psi_{+,\bm k'} \hat b_{-\bm k'+\bm q})\bigg]. \nonumber
    \label{eq:Shraiman-Siggia}
\end{align}
Here we have again used the mode decompositions in Eqs. \eqref{mode1} and \eqref{mode2}, and $\varepsilon_\k$ is the dispersion generated by the nearest and next-nearest neighbor hoppings. A detailed derivation of the Shraiman-Siggia interaction term is given in the appendix.

\begin{figure*}
\includegraphics[width=\linewidth]{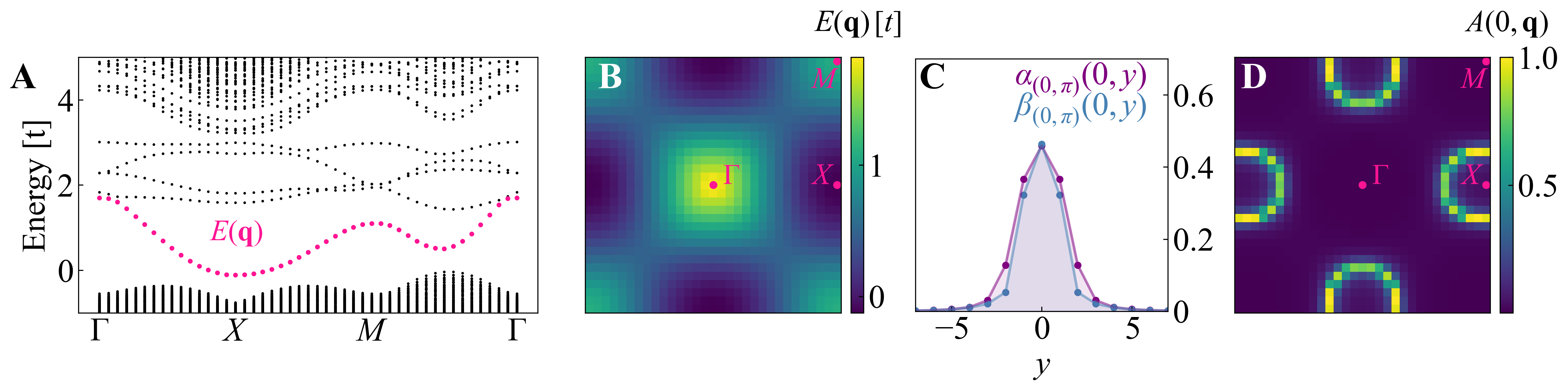} 
\caption{\textbf{A:} 2-particle spinon-chargon spectrum along a high-symmetry cut through the Brillouin zone, obtained with $t=1$, $t'=-0.2$, $g=2.6$, $\chi=1$, $v=2$, and $m=0.1$. \textbf{B:} Plot of the lowest bound-state band dispersion $E(\bm q)$. \textbf{C:} Absolute value of the $\bm q=(\pi,0)$ real-space bound state wavefunctions $\alpha(\bm r)$ and $\beta(\bm r)$ along $(0,y)$.  \textbf{D:} Proxy for the electron spectral function defined in Eq. \eqref{proxy}, where $\epsilon = 0.05$ and electrons occupying $15\%$ of the BZ area.} 
\label{fig:side}
\end{figure*}

\subsection{Total Hamiltonian and bound state wave functions}
The Hamiltonian defining our spinon-chargon bound state problem is $\hat H = \hat H_0 + \hat H_C + \hat H_{\psi z}$, where $\hat H_0$ is the non-interacting part:
\begin{eqnarray}
\hat H_0 & = & \sum_\q \omega_\q (\hat a_\q^\dagger \hat a_\q + \hat b^\dagger_\q \hat b_\q) \\
 & & + \sum_\k \varepsilon_\k \hat \psi^\dagger_\k \hat \psi_\k + g \sum_\k \hat \psi^\dagger_\k \sigma^z \hat \psi_{\k+\Q}\, , \nonumber
\end{eqnarray}
with $\Q = (\pi,\pi)$ the AFM wave vector. To numerically obtain the bound states we work in the folded or magnetic Brillouin zone (MBZ) with reciprocal vectors $\Q$ and $(-\pi,\pi)$, as the staggered potential couples wavevectors $\k$ and $\k+\Q$. 

For every $\q$ in the MBZ, there are two types of bound state wave functions which are eigenstates of the translation operator. To see this, first note that translation acts on the Fourier components of the chargon fields as
\begin{eqnarray}
\hat \psi_{+,\k} & \rightarrow & -e^{-ik_{x/y}} \hat \psi_{-,\k}, \\
\hat \psi_{-,\k} & \rightarrow & e^{-ik_{x/y}} \hat \psi_{+,\k}
\end{eqnarray}
On the Fourier components of the spinon field translation acts as $z_\q \rightarrow e^{-iq_{x/y}}\mathrm{i}\sigma^y (z_{-\q})^*$, which implies that
\begin{equation}
\hat b_\q + \mathrm{i}\sigma^y \hat a^\dagger_{-\q} \rightarrow e^{-iq_{x/y}}\mathrm{i}\sigma^y(\hat b^\dagger_{-\q} + \mathrm{i}\sigma^y \hat a_{\q}).
\end{equation}
This shows that the spinon mode operators transform as
\begin{eqnarray}
\hat a_{\q,\sigma} & \rightarrow & e^{-iq_{x/y}} \hat b_{\q,\sigma}, \\
\hat b_{\q,\sigma} & \rightarrow & -e^{-iq_{x/y}} \hat a_{\q,\sigma}\,,
\end{eqnarray}
where $\sigma$ is the spin index. The first type of bound state which is an eigenstate under this action of translation is given by
\begin{align} 
    \hat f^\dagger_{\q,n,\sigma} = &\sum_{\bm k}\alpha_{\bm q}^n(\bm k)(\hat \psi_{+,\bm k}^\dagger \hat b_{-\bm k+\bm q,\sigma}^\dagger+ \hat \psi_{-,\bm k}^\dagger \hat a_{-\bm k+\bm q,\sigma}^\dagger) \label{MBZ1} \\
    &+ \beta_{\bm q}^n(\bm k)(\hat \psi_{+,\bm k+\bm Q}^\dagger\hat b_{-\bm k+\bm q,\sigma}^\dagger -\hat\psi_{-,\bm k+\bm Q}^\dagger \hat a_{-\bm k+\bm q,\sigma}^\dagger). \nonumber 
\end{align}
Here we use the index $n$ to distinguish different eigenstates. Under a single-site translation this bound state picks up a phase $e^{iq_{x/y}}$, which shows that it indeed has momentum $\q$. The second type of bound state takes the form
\begin{align}
    \hat f^\dagger_{\q+\Q,n,\sigma} = & \sum_{\bm k}\gamma_{\bm q}^n(\bm k)(\hat\psi_{+,\bm k}^\dagger \hat b_{-\bm k+\bm q,\sigma}^\dagger-\hat\psi_{-,\bm k}^\dagger \hat a_{-\bm k+\bm q,\sigma}^\dagger) \nonumber  \\
    &+ \delta_{\bm q}^n(\bm k)(\hat \psi_{+,\bm k+\bm Q}^\dagger \hat b_{-\bm k+\bm q,\sigma}^\dagger + \hat \psi_{-,\bm k+\bm Q}^\dagger \hat a_{-\bm k+\bm q,\sigma}^\dagger). \label{MBZ2}
\end{align}
This state picks up a phase $-e^{iq_{x/y}}$ under translations, showing it has momentum $\q+\Q$, hence our notation.

After obtaining the eigenstates for wavevectors $\q$ in the first MBZ it is trivial to unfold the spectrum: if the states have eigenvalue $e^{iq_x}$ under single-site translation along $x$ we assign them to momentum point $\q$ the first MBZ, and if they have eigenvalue $-e^{iq_x}$ we assign them to the momentum point $\q+\Q$ in the second MBZ. 

If we call the momentum point in the second MBZ $\q'=\q+\Q$, then we can rewrite Eq. \eqref{MBZ2} as
\begin{align}
    \hat f^\dagger_{\q',n} = & \sum_{\bm k}\gamma_{\bm q' +\Q}^n(\bm k + \Q)(\hat \psi_{+,\bm k + \Q}^\dagger \hat b_{-\bm k+\bm q' }^\dagger-\hat \psi_{-,\bm k + \Q }^\dagger \hat a_{-\bm k+\bm q' }^\dagger) \nonumber\\
    &+ \delta_{\bm q' + \Q}^n(\bm k + \Q)(\hat\psi_{+,\bm k}^\dagger \hat b_{-\bm k+\bm q'}^\dagger +\hat \psi_{-,\bm k}^\dagger \hat a_{-\bm k+\bm q'}^\dagger). \nonumber
\end{align}
Here we have left the spin index implicit. This bound state is of exactly the same form as in Eq.~\eqref{MBZ1}, provided that we define $\alpha^n_{\q'}(\k) = \delta^n_{\q'+\Q}(\k+\Q)$ and $\beta^n_{\q'}(\k) = \gamma^n_{\q'+\Q}(\k+\Q)$ when $\q'$ is in the second MBZ. From now on we will use this definition and write all bound states as in Eq.~\eqref{MBZ1}, with $\q$ running over the entire Brillouin zone of the square lattice.

\subsection{Numerical results}

In Fig. \ref{fig:side} we show 2-particle exact diagonalization results for a $30\times 30$ system, obtained with $t=1$, $t'=-0.2$, $g=2.6$, $\chi=1$, $v=2$, and $m=0.1$. Note that $m$ should in principle be determined self-consistently from the other parameters in the theory. Here, however, we will simply treat it as a phenomenological parameter. With these parameters we can estimate the anti-ferromagnetic correlation length as half the spinon correlation length: $\xi_{AFM} \sim v/2m = 10$ lattice sites. We have also used $\tilde{t}_{\r\r'}=t_{\r\r'}$.

In Fig. \ref{fig:side} \textbf{A} we show the spinon-chargon spectrum $E(\q)$ along a high-symmetry cut through the Brillouin zone. Among the positive-energy states we recognize five bound state bands which have split off from the two-particle continuum. The lowest of these five bands, highlighted in pink, has minima at the $X$-points, i.e. at $(\pi,0)$ and $(0,\pi)$, and a bandwidth of $\sim 1.7t$. In Fig. \ref{fig:side} \textbf{B} we show a colorplot of the energy of this band throughout the entire Brillouin zone. 

To get an idea of the spatial extent of the bound states in Fig. \ref{fig:side} \textbf{C} we plot the real-space wave function coefficients $|\alpha_\q(\r)|$ and $|\beta_\q(\r)|$ of the lowest bound-state band at $\q=(0,\pi)$. We see that the linear size of the lowest-energy bound states is $\sim 2-3$ lattice sites.

The bound state wave functions will introduce non-trivial matrix elements in the electron spectral function. We can for example calculate following proxy for the zero-frequency spectral function:
\begin{equation}
A(0,\q) = \frac{1}{\pi}\frac{\epsilon}{(E(\q) - \mu)^2 + \epsilon^2} |\langle 0 |\hat c_\q \hat f^\dagger_\q|0\rangle|^2\,, \label{proxy}
\end{equation}
with $E(\q)$ and $f^\dagger_\q$ the energy and creation operators of the lowest bound-state band. In Fig. \ref{fig:side} \textbf{D} we plot $A(0,\q)$ normalized to a maximal value of one, obtained with $x=15\%$ and $\epsilon=0.05$, as a function of $\q$. We see that the matrix elements cause the spectral weight of the electron pockets centered at the $X$-points to vary along the Fermi surface. In particular, the spectral weight on the sides of the pockets facing the $\Gamma$ point is smaller than at the sides facing the $M$ point.

\begin{figure}
\centering
\includegraphics[width=0.8\linewidth]{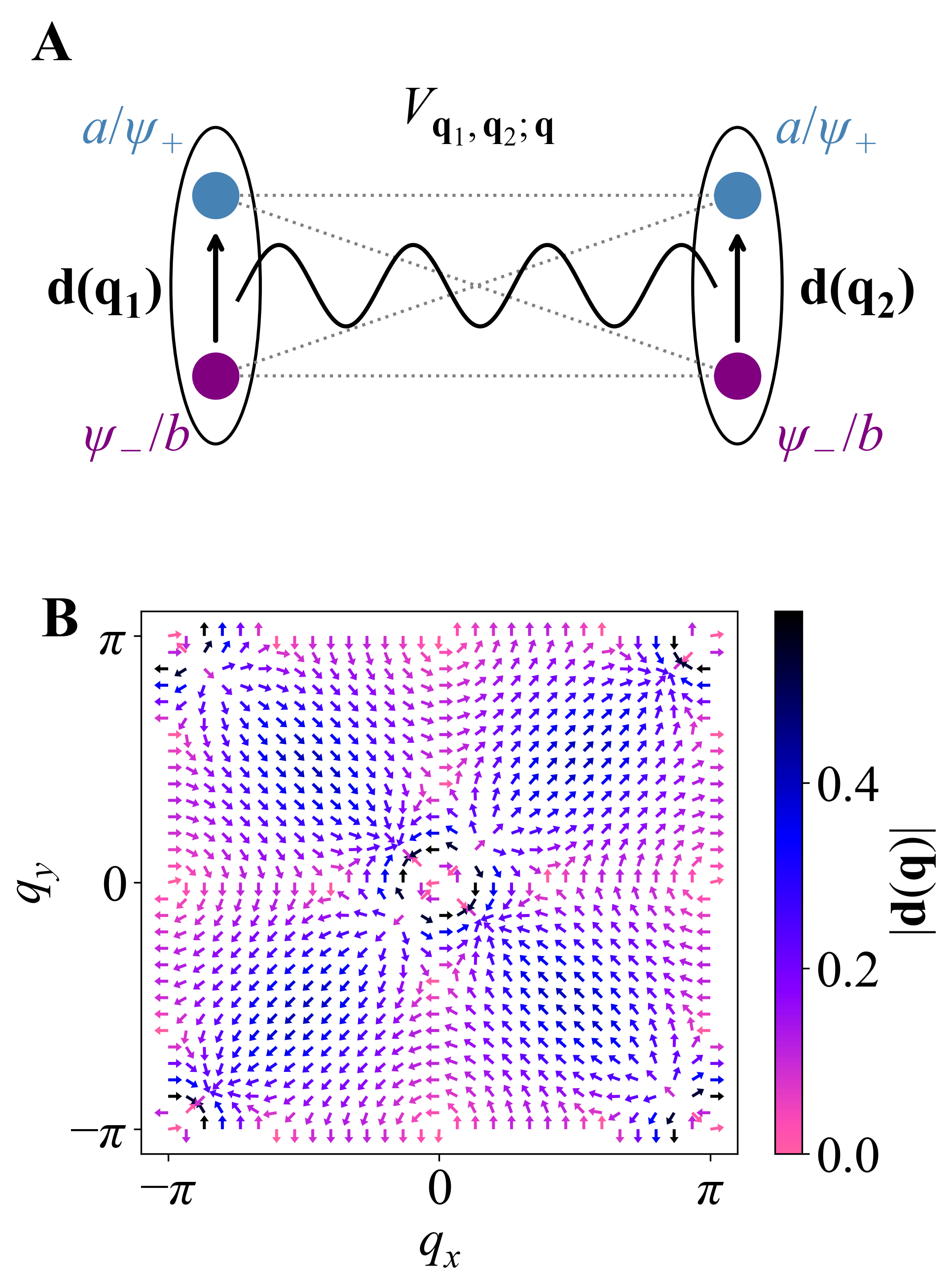} 
\caption{\textbf{A:} Schematic of the effective interaction between two spinon-chargon bound states with a finite effective dipole moment. The dipolar interaction is generated by attractive (repulsive) interactions between opposite (equal) gauge charges. \textbf{B:} The effective dipole moment distribution of the bound states in the BZ. } 
\label{fig:dipole_dipole}
\end{figure}

\section{Dipolar interaction}

Having obtained the spinon-chargon bound states, we now determine the interaction between a pair of such bound states. If we write this interaction as
\begin{equation}
    \hat{V} = \frac{1}{2N}\sum_{\bm q_1 \bm q_2 \bm q}\sum_{\sigma,\sigma'}V_{\bm q_1,\bm q_2;\bm q} : \hat{f}_{\bm q_1+\bm q,\sigma}^\dagger \hat{f}_{\bm q_1,\sigma} \hat{f}_{\bm q_2-\bm q,\sigma'}^\dagger \hat{f}_{\bm q_2,\sigma'}:\,,
\end{equation}
with $f^\dagger_{\q,\sigma}$ the creation operator of a bound state with momentum $\q$ and spin $\sigma$ in the lowest-energy band, then our goal is to determine $V_{\q_1,\q_2;\q}$. For this we first invert Eqs. \eqref{MBZ1} and \eqref{MBZ2} to obtain (with spin indices implicit)
\begin{equation}
\hat \psi^\dagger_{+,\k}\hat b^\dagger_{-\k+\q} = \sum_n \alpha^n_\q(\k) \hat f^\dagger_{\q,n} + \beta^n_{\q+\Q}(\k+\Q) \hat f^\dagger_{\q+\Q,n}.
\end{equation}
In this expression we have used that the wave functions are real. We now project in the lowest bound-state band by truncating the sum over $n$, and only keeping the contribution from the band of interest:
\begin{equation}
\hat \psi^\dagger_{+,\k}\hat b^\dagger_{-\k+\q} \rightarrow \alpha_\q(\k) \hat f^\dagger_{\q} + \beta_{\q+\Q}(\k+\Q) \hat f^\dagger_{\q+\Q}.  \label{proj1}
\end{equation}
For the other bilinear we similarly find
\begin{equation}
\hat \psi^\dagger_{-,\k} \hat a^\dagger_{-\k+\q} \rightarrow \alpha_\q(\k) \hat f^\dagger_{\q} - \beta_{\q+\Q}(\k+\Q) \hat f^\dagger_{\q+\Q}.  \label{proj2}
\end{equation}
As sketched in Fig.~\ref{fig:dipole_dipole} \textbf{A}, two bound states interact via a sum of four Coulomb interactions: two repulsive interactions between spinons or chargons with the same gauge charge, and two attractive interactions between opposite gauge charges. After acting with the Coulomb interactions on the state $\hat f^\dagger_{\q_1} \hat f^\dagger_{\q_2}|0\rangle$, we can project it back in the lowest bound-state band by using Eqs. \eqref{proj1} and \eqref{proj2}. This produces the result
\begin{equation}
    V_{\bm q_1,\bm q_2;\bm q} = V_{\bm q + \Q} \lambda(\bm q_1,\bm q_1+\bm q) \lambda(\bm q_2,\bm q_2-\bm q).
    \label{eq:bound_state_interaction}
\end{equation}
The detailed derivation is given in the appendix. Note that the interaction potential is shifted by $\Q$, which implies that it now diverges at the AFM wavevector. The form factors in the bound state interaction are given by
\begin{align*}
\lambda(\bm q_1,\bm q_1+\bm q) &= 2\sum_{\bm k} \bigg[(\alpha_{\bm q_1+\bm q}(\bm k+\bm q)-\alpha_{\bm q_1+\bm q}(\bm k+\bm Q))\beta_{\bm q_1}(\bm k) \\
&+(\beta_{\bm q_1+\bm q}(\bm k+\bm q)-\beta_{\bm q_1+\bm q}(\bm k+\bm Q))\alpha_{\bm q_1}(\bm k)\bigg].
    \label{eq:form_factor}
\end{align*}
From this expression we see that $\lambda(\q_1,\q_1+\Q)=0$. Expanding around $\bm q=\bm Q$, at lowest order the interaction takes on a dipolar form
\begin{equation}
    V_{\bm q_1,\bm q_2;\bm q} = 16\pi m \frac{(\bm q-\bm Q)\cdot \bm d(\bm q_1) \, (\bm q-\bm Q)\cdot\bm d(\bm q_2)}{(\bm q-\bm Q)^2},
    \label{eq:dipole_dipole_interaction}
\end{equation}
where $\d(\q_1)$ is an effective dipole moment, given in terms of the bound state wavefunctions by
\begin{align}
    \bm d(\bm q_1)=2\sum_{\bm k} \bigg[&\nabla_{\bm k} \alpha_{\bm q_1+\bm Q}(\bm k+\bm Q) \beta_{\bm q_1}(\bm k) \\
    & \hspace{0.5 cm}+ \bm \nabla_{\bm k}\beta_{\bm q_1+\bm Q}(\bm k+\bm Q) \alpha_{\bm q_1}(\bm k)\bigg]. \nonumber
\end{align}
The effective dipole moment is plotted in Fig.~\ref{fig:dipole_dipole} \textbf{B}. 

The $\Q$-shift in the interaction potential and the expression for the form factors follow directly from the sign structure of the bound state wave function in Eq. \eqref{MBZ1}, which is in turn fixed by the projective action of translation on the spinons and chargons. The dipolar interaction obtained in this section is therefore very general, and does not depend on energetic details.

\section{$d_{x^2-y^2}$-wave superconductivity}
As a final step we consider a finite density of spinon-chargon bound states doped in the half-filled state. We take the bound states to be rigid and given by the solution of the two-particle problem -- an approximation that should become exact in the dilute limit. As the bound states are fermions, they will form electron pockets centered at the $X$-points in the Brillouin zone.

We now consider the dipolar interaction derived in the previous section in the Cooper channel:
\begin{equation}
    V_{\bm q_1,-\bm q_1;\bm q} = V_{\bm q + \Q} \lambda(\bm q_1,\bm q_1+\bm q) \lambda(-\bm q_1,-\bm q_1-\bm q).
\end{equation}
From the time-reversal or reflection symmetry we know that $\lambda(-\q_1,-\q_1-\q) = \pm \lambda(\q_1,\q_1+\q)$. It is always possible to choose a gauge for the bound state wave functions such that we get this equality with the plus sign. Other gauge choices merely implement a similarity transformation on the interaction in the Cooper channel, and hence will not affect our conclusions below. We thus find
\begin{equation}
    V_{\bm q_1,-\bm q_1;\bm q} = V_{\bm q + \Q} |\lambda(\bm q_1,\bm q_1+\bm q)|^2 \,,
\end{equation}
which shows that the dipolar interaction is repulsive. As $V_{\q+\Q}$ diverges at $\q=\Q$, and $|\lambda(\q_1,\q_1+\q)| \leq 1$, the dipolar interaction will be peaked at $\q=\Q$ as long as the effective dipole moment $\d(\q_1)$ introduced in the previous section is not too small.
 
\begin{figure}
    \centering 
    \includegraphics[width=0.8\linewidth]{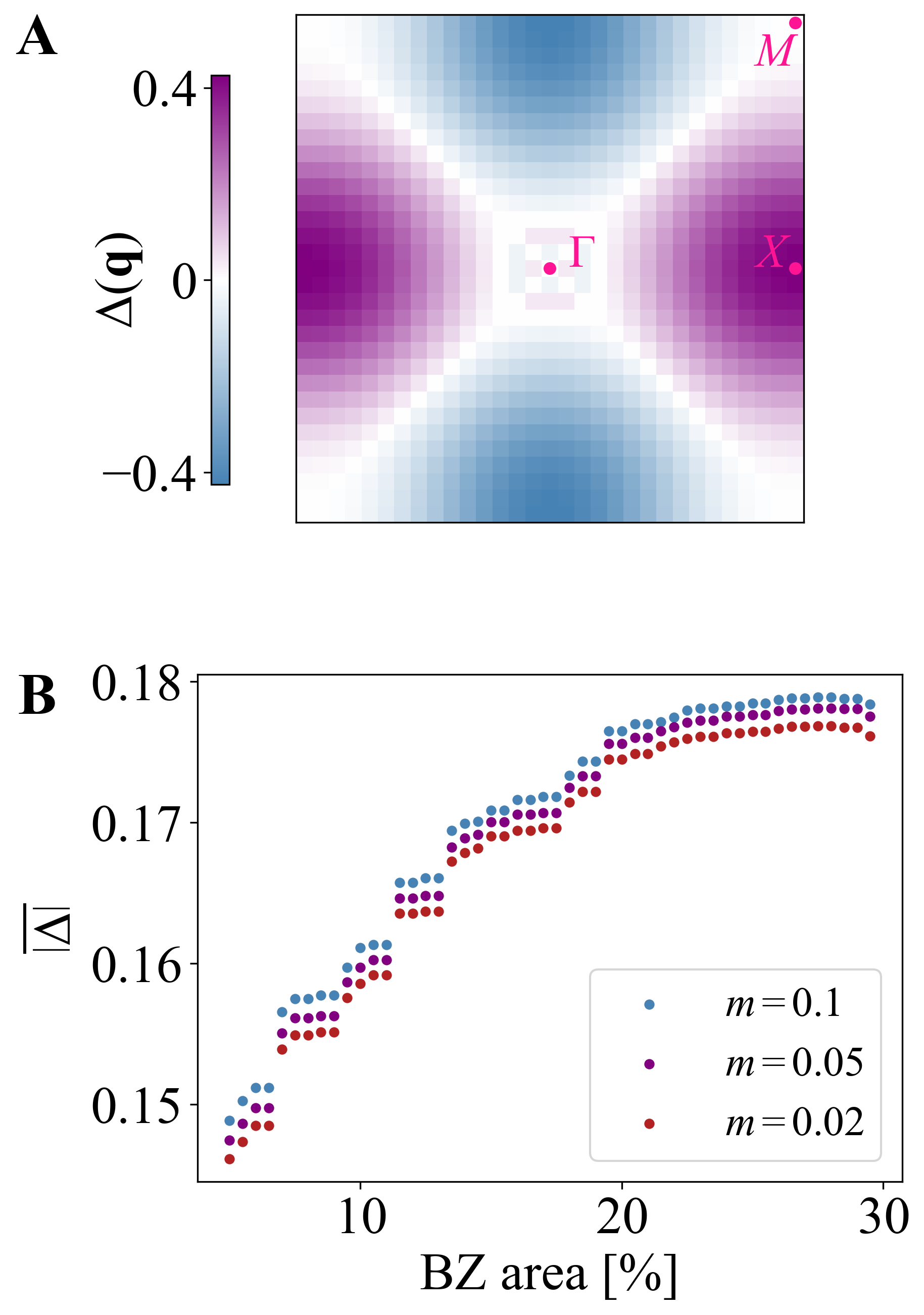}
    \caption{Superconducting order parameter $\Delta(\bm q)$ obtained by solving the BCS self-consistency equation at temperature $T=0.01 t$ (= $35$ K for $t=0.3\,$eV). \textbf{A:} Order parameter $\Delta(\bm q)$ with $m=0.1$ and electrons occupying $15\%$ of the BZ area. \textbf{B:} The BZ-averaged order parameter as a function of the fraction of the BZ area contained in the electron pockets for different values of $m$.} 
    \label{fig:Superconductivity} 
\end{figure}

A repulsive interaction peaked at $\Q$ is exactly what is needed to get $d_{x^2-y^2}$ superconductivity \cite{Hirsch1985,Emery1986,Scalapino1986,Miyake1986}. By numerically solving the self-consistent BCS equation we indeed find a $d$-wave gap, as shown in Fig. \ref{fig:Superconductivity} \textbf{A}. This result was obtained at a temperature $T=0.01 t$ (which corresponds to $35$ K if we use $t=0.3$ eV), and with the electron pockets occupying $15\%$ of the Brillouin zone area.

In Fig. \ref{fig:Superconductivity} \textbf{B} we show the average gap magnitude as a function of the fraction of the Brillouin zone contained in the electron pockets, and for different values of $m$. As expected, the average gap increases with the area of the electron pockets. Note that this area is not necessarily tied to the filling, as the electron pockets are sometimes accompanied by hole pockets centered at $(\pm \pi/2,\pm \pi/2)$, which we have not taken into account in this work. But even without hole pockets, where an electron doping density $n=0.15$ would lead to electron pockets occupying $7.5 \%$ of the Brillouin zone, we find a sizable superconducting gap. We also see that for $m$ in the range $[0.02-0.1]$ the gap depends only weakly on the spinon mass. This is due to two competing effects: a smaller $m$ leads to a weaker Coulomb interaction, which makes the overall scale of the dipolar interaction smaller, but also leads to a larger effective dipole moment $\d(\q_1)$ as the bound states become more spread out in real space.

\section{Discussion}
To arrive at our results we have assumed that the half-filled state with gapped spinons and chargons is described by a deconfined compact U(1) gauge theory. This cannot really be true, however, and at the largest length scales the U(1) gauge theory must confine. Nevertheless, this does not pose a fundamental problem. First, it is possible that with finite doping, the deconfined phase becomes stable. But even if this is not the case (for example, because the doped electrons enter as gauge neutral bound states), our results should go through as long as the deconfined U(1) gauge theory description applies up to a length scale which is larger than the size of the spinon-chargon bound states and the Cooper pairs. After all, our numerical results were obtained on a $30\times 30$ system, so if the FL$^*$ description holds up to this scale, our analysis is valid. 

Finally, an interesting question that we leave for future work is to obtain an estimate for $T_c$. We emphasize that the temperature-dependence of the superconducting gap will not follow the standard BCS behaviour, as the pair-forming interaction itself is temperature dependent (for example, through the dependence of $m$ on temperature). It would also be interesting to explore whether similar ideas as the ones used here are applicable in the context of the SU(2) gauge theory for the cuprate pseudogap obtained from the ancilla-layer model \cite{Zhang2020_2,Christos2023,Christos2024,Christos2024_2}.

\section*{Acknowledgements}
We thank Pietro Bonetti and Jutho Haegeman for helpful and stimulating discussions. This research was supported by the European Research Council under the European Union Horizon 2020 Research and Innovation Programme via Grant Agreement No. 101076597-SIESS (M.S. and N.B.) and by a grant from the Simons Foundation (SFI-MPS-NFS-00006741-04) (A.B and N.B.).

\bibliography{bib}

\newpage
\onecolumngrid
\appendix 
\section{Derivation of the Shraiman-Siggia interaction}
In this section, we derive the form of the spinon-chargon Hamiltonian $H_{\psi z}$. We start from the Shraiman-Siggia action
\begin{eqnarray}
S_{\psi z} =\int\mathrm{d}\tau \bigg[&\sum_{\r,\r'} t_{\r\r'} \bar{\psi}_{\r,-} \psi_{\r',+}\, z_{\r'}\mathrm{i}\sigma^y z_\r 
 + \sum_\r \bar{\psi}_{\r,-} \psi_{\r,+}\, z_{\r}\mathrm{i}\sigma^y \partial_\tau z_\r + \text{h.c.}  \bigg].
 \label{eq:AppSSA}
\end{eqnarray}
The hopping is to nearest and next-nearest neighbors, i.e. 
\begin{equation}
    t_{\bm r\bm r'} = \begin{cases}
        t \quad \bm r'=\bm r\pm \e_{x/y}\\
        t' \quad \bm r'=\bm r \pm \e_{x} \pm \e_{y}.
    \end{cases}
\end{equation}
Inserting the Fourier transforms of the fields
\begin{align}
    z_{\r} &= \frac{1}{\sqrt{N}}\sum_{\q} e^{\mathrm{i}\q\r}z_{\q}, \\
    \psi_{\r} &= \frac{1}{\sqrt{N}}\sum_{\q} e^{\mathrm{i}\q\cdot \r} \psi_{\q}.
\end{align} 
The conjugate momentum of this theory is
\begin{equation}
    \pi_{\bm q} = \frac{\delta\mathcal{L}_{\psi z}}{\delta(\partial_\tau z_{\bm q})} =  \frac{1}{N}\sum_{\bm p\bm p'} \psi_{-,\bm p'}^\dagger\psi_{+,\bm p'-\bm p}z_{\bm p-\bm q}(\mathrm{i}\sigma^y).
\end{equation} 
Performing the Legendre transformation $H_{\psi z} = \sum_\q \pi_\q \partial_\tau z_\q - \mathcal{L}_{\psi z}$, the Hamiltonian reads 
\begin{align} 
    H_{\psi z} = \frac{1}{N^2} \sum_{\bm r}\sum_{\bm q_1\bm q_2\bm q_3\bm q_4}e^{\mathrm{i}(-\bm q_1+\bm q_2+\bm q_3+\bm q_4)\bm r} \varepsilon_{\bm q_1-\bm q_4} \psi_{\q_1,-}^\dagger\psi_{\q_2,+}z_{\q_3}\mathrm{i}\sigma^yz_{\q_4} + \text{h.c.},
\end{align}
where $\varepsilon(\q) = -2t\bigl(\cos(q_xa)+\cos(q_ya)\bigr)-4t'\cos(q_xa)\cos(q_ya))$ is the usual dispersion generated by nearest and next-nearest neighbor hoppings. Inserting the mode expansion of the spinon field $z_{\bm q} = \frac{1}{\sqrt{2\omega_{\q}\chi}}(b_{\q}+\mathrm{i}\sigma^y a_{-\q}^\dagger)$, we have 
\begin{align}
    \hat H_{\psi z} &= \frac{1}{N^2} \sum_{\bm r}\sum_{\bm q_1\bm q_2\bm q_3\bm q_4}
    e^{\mathrm{i}(-\bm q_1+\bm q_2+\bm q_3+\bm q_4)\bm r} 
    \frac{\varepsilon_{\bm q_1-\bm q_4}}{2\chi\sqrt{\omega_{\bm q_3}\omega_{\bm q_4}}} \notag \\
    &\quad \times \Big(
        \hat\psi_{\q_1,-}^\dagger \hat\psi_{\q_2,+} \hat a_{-\q_4}^\dagger \hat b_{\q_3} 
        - \hat \psi_{\q_1,-}^\dagger \hat \psi_{\q_2,+} \hat a_{-\q_3}^\dagger  \hat b_{\q_4} 
    \Big) + \text{h.c.} \notag 
\end{align}
This becomes
\begin{align}
    \hat H_{\psi z} &= \frac{1}{N} \sum_{\bm k\bm k'\bm q} \frac{\varepsilon_{\bm k'+\bm k-\bm q}-\varepsilon_{\bm q}}{2\chi \sqrt{\omega_{-\bm k+\bm q}\omega_{-\bm k'+\bm q}}} \biggl(\hat \psi_{\k,+}^\dagger \hat b_{-\bm k+\bm q}^\dagger\hat\psi_{\bm k',-}\hat a_{-\bm k'+\bm q} + \hat\psi_{\k,-}^\dagger \hat a_{-\bm k+\bm q}^\dagger \hat\psi_{\bm k',+} \hat b_{-\bm k'+\bm q}\biggr).
\end{align}

\section{Details of the numerical simulations}

\subsection{Coulomb interaction}
The Coulomb potential obtained via the spinon polarization is 
\begin{equation}
V_\q = \Pi(0,v\q)^{-1} = \left[\frac{v^2\q^2+4m^2}{8\pi v^3|\q| } \arctan\frac{v|\q|}{2m} - \frac{m}{4\pi v^2}\right]^{-1}\,.
\label{eq:Coulomb_momentum}
\end{equation}
As this expression was obtained via a continuum approximation for the spinons, we need to restore the periodicity under shifts by reciprocal lattice vectors. We do this by replacing $|\bm q| \to 2\sqrt{\sin^2(q_x/2)+\sin^2(q_y/2)}$. We find that in real space this procedure leads to a very large on-site interaction. To reduce this on-site interaction we add an additional strictly on-site interaction term by hand. The precise value of this shift should not matter, as long as it removes the large on-site interaction which results from the replacement $|\bm q| \to 2\sqrt{\sin^2(q_x/2)+\sin^2(q_y/2)}$. This introduces some arbitrariness, and we simply take the real-space interaction to satisfy
\begin{equation}
    \tilde V_\r = \begin{cases}
        V_\r &  \r \neq 0, \\
        V_{(a,0)} & \r=0,
    \end{cases}
\end{equation}
The Fourier transform of $\tilde V_\r$ gives the momentum-space potential $\tilde V_\q$ that we use in our numerics. 
\subsection{Brillouin zone unfolding}
To solve the spinon-chargon bound state problem numerically, we work in the folded or first magnetic Brillouin zone (MBZ). This means we obtain two types of bound state wavefunctions that have momentum eigenvalue $\q$ in the first MBZ or momentum eigenvalue $\q+\Q$ in the second MBZ. To unfold the spectrum, i.e. associate every bound state with the correct momentum in the full Brillouin zone, we replace the Hamiltonian
\begin{equation}
    H \mapsto (H-a\mathbbm{1}) \sigma^z \otimes \tau^x \otimes \mathbbm{1}_{N\times N},
    \label{eq:liftedH}
\end{equation}
where $\sigma^z$ acts on $(\psi_\q \, \psi_{\q+\Q})$ and $\tau^x$ flips the gauge charge. The energy of the bound states with momentum $\q$ in the first MBZ and $\q+\Q$ in the second MBZ are then shifted by $-a$ and $+a$ respectively and states can easily be associated with the correct momentum.
\subsection{Superconductivity}
To obtain the superconducting order parameter at temperature $T$, we numerically solve the BCS self-consistency equation
\begin{equation}
    \Delta_{\q_1}(T)=-\sum_{\q}V_{\q_1,-\q_1;\q}\frac{\Delta_{\q_1+\q}(T)}{2\sqrt{[E(\q_1+\q)]^2+|\Delta_{\q_1+\q}(T)|^2}}\tanh \Biggl[\frac{\sqrt{[E(\q_1+\q)]^2+|\Delta_{\q_1+\q}(T)|^2}}{2k_BT} \Biggr],
\end{equation}
where $E(\q)$ is the bound state dispersion (see Fig.~\ref{fig:side} \textbf{A}) and $V_{\q_1,-\q_1;\q}$ are the dipolar interaction matrix elements that we derive from the bound state wavefunctions in the next section.

\section{Derivation of dipolar interaction}
\begin{figure}
\centering
\includegraphics[width=0.7\linewidth]{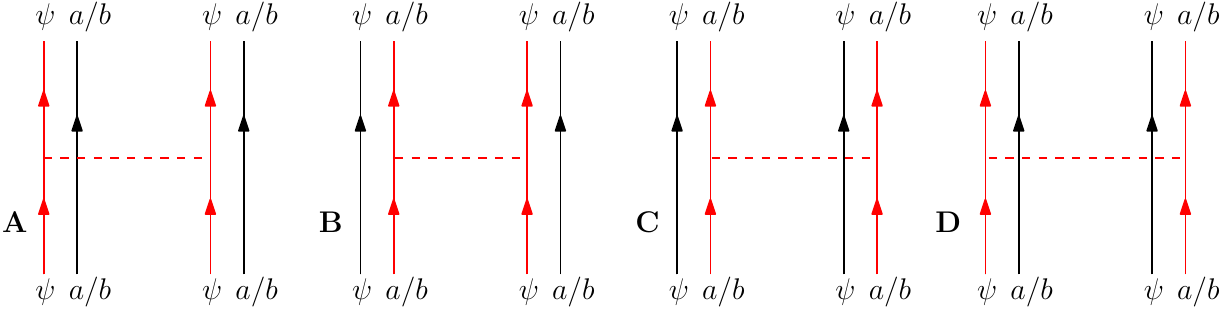} 
\caption{Four scattering processes contributing to the effective dipolar interaction between spinon-chargon pairs. The dashed line represents the Coulomb interaction.}
\label{fig:ints}
\end{figure}

From the form of the Coulomb interaction between opposite gauge charges we derive the effective interaction between bound states. The fermionic creation operators of bound states are defined as
\begin{align}
     \hat f_{n,\bm q}^\dagger &= \sum_{\bm k} \alpha_{\bm q}^n(\bm k)(\hat \psi_{+,\bm k}^\dagger \hat b_{-\bm k+\bm q}^\dagger + \hat\psi_{-,\bm k}^\dagger \hat a_{-\bm k+\bm q}^\dagger)+\beta_{\bm q}^n(\bm k)(\hat \psi_{+,\bm k+\bm Q}^\dagger \hat b_{-\bm k+\bm q}^\dagger - \hat\psi_{-,\bm k+\bm Q}^\dagger \hat a_{-\bm k+\bm q}^\dagger), \qquad \\
     \hat f_{n,\bm q+\bm Q}^\dagger &= \sum_{\bm k} \gamma_{\bm q}^n(\bm k)(\hat\psi_{+,\bm k}^\dagger \hat b_{-\bm k+\bm q}^\dagger - \hat \psi_{-,\bm k}^\dagger \hat a_{-\bm k+\bm q}^\dagger)+\delta_{\bm q}^n(\bm k)(\hat\psi_{+,\bm k+\bm Q}^\dagger \hat b_{-\bm k+\bm q}^\dagger + \hat\psi_{-,\bm k+\bm Q}^\dagger \hat a_{-\bm k+\bm q}^\dagger)\,,
\end{align} 
where the spin indices are implicit. The operator $\hat f_{n,\q}^\dagger$ creates a bound state with momentum $\q$ in the first MBZ and $\hat f_{n,\q+\Q}^\dagger$ creates a bound state in the second MBZ. As explained in the main text, by comparing the two operators we can identify
\begin{align}
    \gamma_{\bm q}^n(\bm k+\bm Q) &= \beta_{\bm q + \bm Q}^n(\bm k), \\
    \delta_{\bm q}^n(\bm k+\bm Q) &= \alpha_{\bm q + \bm Q}^n(\bm k).
\end{align} 
We expand the spinon-chargon pairs with total momentum $\bm q$ as 
\begin{align}
     \hat \psi_{+,\bm k}^\dagger \hat b_{-\bm k+\bm q}^\dagger = \sum_n ([\alpha_{\bm q}^n(\bm k)]^* \hat f_{n,\bm q}^\dagger+[\beta_{\bm q+\bm Q}^n(\bm k+\bm Q)]^*\hat f_{n,\bm q+\bm Q}^\dagger), \\ 
     \hat\psi_{-,\bm k}^\dagger \hat a_{-\bm k+\bm q}^\dagger = \sum_n ([\alpha_{\bm q}^n(\bm k)]^* \hat f_{n,\bm q}^\dagger-[\beta_{\bm q+\bm Q}^n(\bm k+\bm Q)]^*\hat f_{n,\bm q+\bm Q}^\dagger),
\end{align} 
where we used orthonormality of the wavefunctions 
\begin{align}
    & 2 \sum_n \alpha^{n*}_\q(\k) \alpha^{n}_\q(\k') + \beta^{n*}_\q(\k) \beta^{n}_\q(\k') = \delta_{\k,\k'}, \\
    & 2 \sum_n \gamma^{n*}_\q(\k) \gamma^{n}_\q(\k') + \delta^{n*}_\q(\k) \delta^{n}_\q(\k') = \delta_{\k,\k'}.
\end{align}
To project the interaction in the lowest-energy bound state band we make the following approximation
\begin{align}
    \hat \psi_{+,\bm k}^\dagger \hat b_{-\bm k+\bm q}^\dagger &\approx \bigl(\alpha_{\bm q}^*(\bm k) \hat f_{\bm q}^\dagger+\beta_{\bm q+\bm Q}^*(\bm k+\bm Q)\hat f_{\bm q+\bm Q}^\dagger\bigl),\label{eq:bandapprox}  \\
     \hat\psi_{-,\bm k}^\dagger \hat a_{-\bm k+\bm q}^\dagger &\approx \bigl(\alpha_{\bm q}^*(\bm k) \hat f_{\bm q}^\dagger-\beta_{\bm q+\bm Q}^*(\bm k+\bm Q)\hat f_{\bm q+\bm Q}^\dagger\bigr). 
\end{align}
After the band projection we can drop the index $n$, and the effective interaction obtained from the sum of four Coulomb interactions (see Fig.~\ref{fig:side} \textbf{A}) can be written as 
\begin{equation}
    \hat V = \frac{1}{2N}\sum_{\bm q_1 \bm q_2 \bm q}\sum_{\sigma,\sigma'} V_{\bm q_1,\bm q_2;\bm q} : \hat f_{\bm q_1+\bm q,\sigma}^\dagger \hat f_{\bm q_1,\sigma} \hat f_{\bm q_2-\bm q,\sigma'}^\dagger \hat f_{\bm q_2,\sigma'}:\,,
\end{equation} 
where the matrix elements $V_{\bm q_1,\bm q_2;\bm q}$ are a product of the Coulomb potential $V_{\bm q}$ and form factors $\lambda$ resulting from the bound state projection: 
\begin{equation}
    V_{\bm q_1\bm q_2;\bm q} = V_{\bm q+\bm Q} \lambda(\bm q_1,\bm q_1+\bm q) \lambda(\bm q_2,\bm q_2-\bm q).
    \label{eq:dipole_interaction}
\end{equation} 
To sketch the derivation of the form factors, we consider interactions of all possible pairs of fermions and spinons 
\begin{align}
    &\text{1. } \hat\psi_+ \hat b \longleftrightarrow \hat \psi_+ \hat b \, , \label{eq:possiblepairs}\\
    &\text{2. } \hat \psi_-\hat a \longleftrightarrow \hat\psi_-\hat a \, , \nonumber\\
    &\text{3. } \hat \psi_+\hat b \longleftrightarrow \hat \psi_-\hat a \, , \nonumber\\
    &\text{4. } \hat \psi_+ \hat b \longleftrightarrow \hat \psi_- \hat a \, . \nonumber
\end{align} 
We ignore the relative momentum between fermions and spinons for now and define the following short-hand notation 
\begin{align}
    A &= \alpha_{\q_1+\q}^* \hat f_{\q_1+\q}^\dagger,  && E = \alpha_{\q_2-\q}^* \hat f_{\q_2-\q}^\dagger,  \\
    B &=  \beta_{\q_1+\q+\Q}^* \hat f_{\q_1+\q+\Q}^\dagger,  &&F = \beta_{\q_2-\q+\Q}^* \hat f_{\q_2-\q}^\dagger,  \nonumber \\
    C &= \alpha_{\q_1} \hat f_{\q_1},   &&G = \alpha_{\q_2} \hat f_{\q_2},  \nonumber \\
    D &= \beta_{\q_1+\Q} \hat f_{\q_1+\Q},   && H = \beta_{\q_2+\Q} \hat f_{\q_2+\Q} \, , \nonumber 
    \label{eq:notation}
\end{align}
with spin indices again implicit. Summing all possibilities of scattering pairs (see eq.\ref{eq:possiblepairs}) and plugging in the expansions in eq.~\ref{eq:bandapprox}, we end up with
\begin{align} 
    (A+B)(C+D)(E+F)(G+H) \\ \nonumber 
    + (A-B)(C-D)(E-F)(G-H) \\ \nonumber
    - (A+B)(C+D)(E-F)(G-H) \\ \nonumber 
    - (A-B)(C-D)(E+F)(G+H), \nonumber 
\end{align}
where the sign of a term is positive for repulsive and negative for attractive interactions. The only terms that survive this sum are 
\begin{align}
    ADEH + BCFG + ADFG + BCEH.
    \label{eq:ABCD}
\end{align} 
 For each of the four terms in eq.\ref{eq:ABCD}, the momentum $\q$ can be transferred in four different channels shown in Fig.\ref{fig:ints} \textbf{A}-\textbf{D}. For example, consider the term $ADEH$ and the particles scattering in the \textbf{A} channel. This term of the interaction is 
\begin{align}
    [ADEH]_{\textbf{A}} &= \sum_{\q_1\q_2\q}\sum_{\k_1\k_2} V_\q \alpha_{\q_1+\q}^*(\k_1+\Q+\q) \beta_{\q_1+\Q}(\k_1)\alpha_{\q_2-\q}^*(\k_2+\Q-\q) \beta_{\q_2+\Q}(\k_2) \hat f_{\q_1+\q}^\dagger \hat f_{\q_1+\Q} \hat f_{\q_2-\q}^\dagger \hat f_{\q_2+\Q} \\
    & = \sum_{\q_1\q_2\q}\sum_{\k_1\k_2} V_{\q+\Q} \alpha_{\q_1+\q}^*(\k_1+\q) \beta_{\q_1}(\k_1) \alpha_{\q_2-\q}^*(\k_2-\q) \beta_{\q_2}(\k_2) \hat f_{\q_1+\q}^\dagger \hat f_{\q_1} \hat f_{\q_2-\q}^\dagger \hat f_{\q_2} \, ,
\end{align}
where we shifted the momenta $\q\to \q+\Q$, $\q_1\to \q_1+\Q$ and $\q_2\to \q_2+\Q$. Note that $2\Q$ is a reciprocal lattice vector. The \textbf{B} channel corresponds to 
\begin{align}
    [ADEH]_{\textbf{B}} &= -\sum_{\q_1\q_2\q}\sum_{\k_1\k_2} V_\q \alpha_{\q_1+\q}^*(\k_1)\beta_{\q_1+\Q}(\k_1+\Q)\alpha_{\q_2-\q}^*(\k_2-\q) \beta_{\q_2+\Q}(\k_2+\Q) \hat f_{\q_1+\q}^\dagger \hat f_{\q_1+\Q} \hat f_{\q_2-\q}^\dagger \hat f_{\q_2+\Q} \\
    &= -\sum_{\q_1\q_2\q}\sum_{\k_1\k_2} V_{\q+\Q} \alpha_{\q_1+\q}^* (\k_1+\Q)\beta_{\q_1}(\k_1)\alpha_{\q_2-\q}^*(\k_2-\q)\beta_{\q_2}(\k_2) \hat f_{\q_1+\q}^\dagger \hat f_{\q_1} \hat f_{\q_2-\q}^\dagger \hat f_{\q_2}.
\end{align}
Working out all other scattering channels in a similar way, we find 
\begin{equation}
    \lambda(\bm q_1,\bm q_1+\bm q) = 2\sum_{\bm k} (\alpha_{\bm q_1+\bm q}^*(\bm k+\bm q)-\alpha_{\bm q_1+\bm q}^*(\bm k+\bm Q))\beta_{\bm q_1}(\bm k) +(\beta_{\bm q_1+\bm q}^*(\bm k+\bm q)-\beta_{\bm q_1+\bm q}^*(\bm k+\bm Q))\alpha_{\bm q_1}(\bm k).
    \label{eq:dipole_form_factor}
\end{equation}

\end{document}